# Real-Time Artificial Intelligence Assistance for Safe Laparoscopic Cholecystectomy: Early-Stage Clinical Evaluation

Pietro Mascagni[1,2], M.D., Ph.D.; Deepak Alapatt[1], M.Sc.; Alfonso Lapergola[3], M.D.; Armine Vardazaryan[4], M.Sc.; Jean-Paul Mazellier[4], Ph.D.; Bernard Dallemagne[5], M.D.; Didier Mutter[3,4], M.D., Ph.D., F.A.C.S, FRSM; Nicolas Padoy[1,4], Ph.D.

1. ICube, University of Strasbourg, CNRS, France
2. Fondazione Policlinico Universitario A. Gemelli IRCCS, Rome, Italy
3. Department of Digestive and Endocrine Surgery, Nouvel Hôpital Civil, Hôpitaux Universitaires de Strasbourg, France
4. IHU-Strasbourg, Institute of Image-Guided Surgery, Strasbourg, France
5. Institute for Research against Digestive Cancer (IRCAD), Strasbourg, France

Artificial intelligence is set to be deployed in operating rooms to improve surgical care. This early-stage clinical evaluation shows the feasibility of concurrently attaining real-time, high-quality predictions from several deep neural networks for endoscopic video analysis deployed for assistance during three laparoscopic cholecystectomies.

Correspondence to: Pietro Mascagni, M.D., Ph.D., pietro.mascagni@ihu-strasbourg.eu
This study was partially supported by French state funds managed by the ANR under grants ANR-20-CHIA-0029-01 (National AI Chair AI4ORSafety) and ANR-10-IAHU-02 (IHU-Strasbourg). Nvidia Corporation provided hardware and technical support for the study.

# Index

Real-Time Artificial Intelligence Assistance for Safe Laparoscopic Cholecystectomy    3## 1. Introduction

Surgery is an indivisible and indispensable part of health care, accounting for about one-third of the global burden of disease (Meara et al., 2015). However, worldwide at least 4.8 billion people lack access to adequate surgical care (Alkire et al., 2015) and surgery is estimated to account for a great part of preventable medical errors (Zegers et al., 2011).

Surgical Data Science (SDS) aims to leverage data and analytics to make surgery more safe, accessible, and efficient (Maier-Hein et al., 2017). Over the last several years, advances in SDS have accompanied the introduction of several artificial intelligence (AI) models that could impact various stakeholders, from surgeons to operating room (OR) staff to hospital administrators. While the integration of AI in healthcare is fast advancing in fields such as radiology, pathology, and gastrointestinal endoscopy, AI assistance in interventional healthcare is lagging (Maier-Hein et al., 2022). Reasons for this are both cultural, with surgery being one of the most conservative components of the healthcare system, and technical, as even modern ORs often don't meet the computational requirements of advanced algorithms such as deep neural networks.

Recently, laparoscopic cholecystectomy (LC), a very common abdominal surgical procedure performed by most surgeons, has been the subject of several AI studies. In particular, both academia and industries have proposed AI-based computer vision (CV) solutions to help prevent bile duct injuries (BDI), a dreaded adverse event of LC leading to a threefold increase in mortality at 1 year (Törnqvist et al., 2012) and an estimated cost of 1 billion dollars per year in the United States alone (Berci et al., 2013). For instance, Madani et al have proposed GoNoGoNet, a deep learning model for intraoperative guidance towards safe and unsafe areas of dissection (Madani et al., 2022), while our group has proposed DeepCVS, a 2-stage neural network to segment hepatocystic anatomy and automatically assess the achievement of the Critical View of Safety (CVS) (Mascagni et al., 2022), a strongly recommended view to prevent the visual perceptual illusion causing major BDIs (Brunt et al., 2020).

This early stage clinical evaluation study aims at demonstrating the feasibility of concurrently deploying in the OR a toolkit of AI-based CV models for real-time cognitive support during laparoscopic cholecystectomy (LC).

## 2. Methods

This early stage clinical evaluation study was designed to avoid AI interference with the standard surgical care, hence it was cleared from the local medical research and ethical committee. The study is reported according to the Developmental and Exploratory Clinical Investigations of Decision support systems driven by Artificial Intelligence (DECIDE-AI) guidelines (Vasey et al., 2022).

### 2.1 Participants

Surgeons working at the Digestive and Endocrine Surgery Department of the Nouvel Hôpital Civil (NHC, Strasbourg, France) who expressed their willingness to participate in the study were recruited. No specific user training was deemed necessary as surgeons did not directly interact with AI systems.

Patients older than 18 years undergoing an LC for benign conditions in November 2021 with recruited surgeons were included in the study following their informed, written consent.

### 2.2 AI systems

The AI systems used in this study are prototypes internally developed by the Computational Analysis and Modeling of Medical Activities (CAMMA) research group (ICube, University of Strasbourg, CNRS, IHU Strasbourg, France).

Such AI systems analyze LC videos to provide cognitive assistance during LC. Specifically, deep neural networks designed for analyzing surgical workflows by recognizing surgical phases (Twinanda et al., 2017), i.e. the steps to successfully complete a procedure, and localizing surgical instruments (Nwoye et al., 2019) could be used to improve awareness and readiness of the surgical team and OR staff alike. Similarly, a deep learning model for semantic segmentation of fine-grained hepatocystic anatomy (Mascagni et al., 2022) could inform surgical trainees on critical anatomical structures. Finally, the same model trained to automatically assess the CVS (Mascagni et al., 2022) could help promote the clinical implementation of this universally recommended safety maneuver.

To enable real-time inference of several computationally expensive deep learning models, each model's performance was optimized individually using TensorRT (NVIDIA Corporation, California, United States of America) and concurrent model execution was used to maximize utilization of graphics processing unit (GPU).

## 2.3 Implementation and safety

The AI-assisted procedures took place in the ORs of the Institute of Image-Guided Surgery (IHU-Strasbourg, France).

A Clara Holoscan (NVIDIA Corporation, California, United States of America), a platform designed to develop AI-enabled medical devices and cleared for experimental clinical use, was installed in the OR. The Clara Holoscan was connected to the endoscopic video system (input) and a screen (output).

During the study, research engineers activated the AI analysis using a specifically developed graphical user interface. Optimized AI models were executed at ~60 frames per second on Clara Holoscan. To show the potential for real-time, intraprocedural assistance, explicit AI predictions were overlayed onto laparoscopic videos.

To guarantee the safety of this early stage clinical evaluation, AI inference was performed on secondary endoscopic video signals and predictions were not feedbacked to surgeons. This set-up guaranteed that any eventual malfunction of the AI system or the experimental setup would have not interfered with the surgical procedure and that AI would have not influenced clinical decisions.

### 2.4 Outcomes

The primary outcome of this study was the rate of malfunctions of the AI systems deployed in OR for assistance during LC. Malfunctions were defined by any technical or non-technical problem affecting the real-time analysis of LC videos. Secondary outcomes included the qualitative assessment of AI predictions overlayed onto laparoscopic videos.

## 3. Results

Deep learning models for surgical phase recognition, instrument and anatomy recognition, and assessment of the CVS were successfully demonstrated during three LCs cases performed by 4 surgeons in the ORs of the Institute of Image-Guided Surgery (IHU-Strasbourg, France) in November 2021.

Included patients underwent LC for symptomatic gallbladder lithiasis. No malfunctions of the AI system occurred during the study. The AI models were successfully deployed during procedures and provided real-time, high-quality predictions. An AI-assisted LC procedure performed on November 25, 2021, was successfully streamed at the 32° Digestive System Surgery Congress (Rome, Italy) and at the 17[th] IFSES World Congress of Endoscopic Surgery (WCES) hosted by the European Association of Endoscopic Surgery (EAES) (Barcelona, Spain) (Figure 1).

Overall, the three AI-assisted LC procedures were successfully completed, there were no postoperative complications, and all patients were discharged in good conditions on the same day.

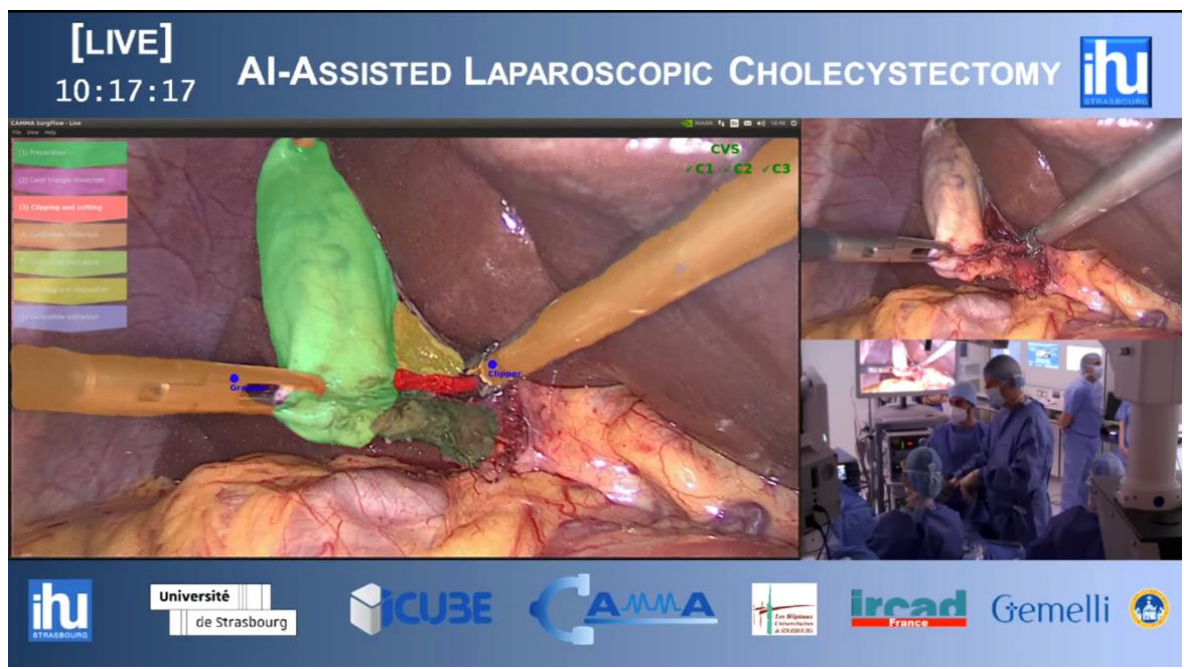

**Figure 1. Live broadcast of AI-assisted LC.** The procedure was performed in the operating room of IHU- Strasbourg on November 25[th], 2021, and live broadcast to the 32° Digestive System Surgery Congress and at the 17[th] IFSES World Congress of Endoscopic Surgery.

## 4. Discussion

Deep learning models for intraoperative cognitive assistance were deployed concurrently and provided real-time, high-quality predictions during three laparoscopic cholecystectomies, one of which was live streamed simultaneously at two international scientific conferences (Figure 1).

To the best of our knowledge, this is the first case series demonstrating the viability of attaining high-quality predictions from a toolkit of computer vision models for real-time assistance during surgery. Computationally expensive deep learning models were optimized on an OR ready platform, concurrently deployed to analyse the same endoscopic video stream, and predictions were overlayed onto laparoscopic videos to sense their quality.

Even though during this proof-of-concept case series surgeons were not exposed to AI predictions, the demonstrated models could be used to improve communication and coordination in the OR, factors known to impact the efficiency and safety of surgical procedures (Graafland et al., 2015), and provide cognitive support to surgeons during critical steps of procedures in the near future (Mascagni & Padoy, 2021).

We believe that by illustrating the technical feasibility and potential clinical value of having live intraoperative feedback from several AI models, this case series could spark research and innovation in AI assistance during interventional healthcare.

Still, several aspects need to be explored before patients and surgeons can benefit from surgical AI. First, methods and tools for secure surgical data sharing need to be developed to gain insights on how to prevent intraoperative adverse events and develop AI solutions robust to variations across hospitals, patient populations, surgical workflows, skills, instrumentation, and acquisition methods. Then, the design of human-machine interfaces should make sure AI tools are ergonomic and their feedback is well-integrated within complex surgical workflows. At this point, pragmatic clinical studies optimized for studying the AI-assisted delivery of care will be necessary to prove clinical value. Finally, the ethical considerations and societal implications of having AI assistance in surgery will have to be addressed so that this novel technology can be effectively deployed for the benefit of patients, surgeons, and healthcare systems worldwide.

## 5. Acknowledgment

The authors would like to thank NVIDIA for their help with the software optimization on its Clara Holoscan platform. They would also like to thank Nicolas Delesse and Oliver Kutter (NVIDIA) for their technical support and Fabio Giannone, Jacques Marescaux, Guido Costamagna, and Benoit Gallix for their clinical advice.

Real-Time Artificial Intelligence Assistance for Safe Laparoscopic Cholecystectomy    11